\documentclass[%
reprint,
superscriptaddress,
 amsmath,amssymb,
 aps,
 prc,
longbibliography,
dvipsnames,
twocolumn
]{revtex4-1}
\usepackage[main=english,german,ngerman]{babel}

\usepackage{xcolor}
\usepackage{graphicx}
\usepackage{dcolumn}
\usepackage{bm}
\usepackage{siunitx}
\usepackage{dsfont}

\usepackage{mathtools }

\usepackage{hyperref}

\usepackage{physics}
\usepackage{bbold}
\usepackage{float}
\usepackage{hyphenat}
\newcommand\figref{Figure~\ref}
\usepackage[utf8]{inputenc}
\usepackage[toc,page]{appendix}

\usepackage{graphicx}
\usepackage{hyperref}
\hypersetup{colorlinks}

\begin{document}
\newcommand{\orcidicon}[1]{\href{https://orcid.org/#1}{\includegraphics[height=\fontcharht\font`\B]{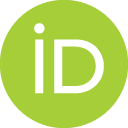}}}
\title{Quantum computing of the $^6$Li nucleus via ordered unitary coupled clusters}

\author{Oriel~Kiss\orcidicon{0000-0001-7461-3342}}
\email{oriel.kiss@cern.ch}
\affiliation{European Organization for Nuclear Research (CERN), Geneva 1211, Switzerland}
\affiliation{Department of Nuclear and Particle Physics, University of Geneva, Geneva 1211, Switzerland}

\author{Michele~Grossi\orcidicon{0000-0003-1718-1314}}
\affiliation{European Organization for Nuclear Research (CERN), Geneva 1211, Switzerland}
 
\author{Pavel~Lougovski\orcidicon{0000-0002-9229-1444}}
\affiliation{AWS Center for Quantum Computing, Seattle, Washington 98121, USA}
 
\author{Federico~Sanchez\orcidicon{0000-0003-0320-3623}}
\affiliation{Department of Nuclear and Particle Physics, University of Geneva, Geneva 1211, Switzerland}

\author{Sofia~Vallecorsa\orcidicon{0000-0002-7003-5765}}
\affiliation{European Organization for Nuclear Research (CERN), Geneva 1211, Switzerland}

\author{Thomas~Papenbrock\orcidicon{0000-0001-8733-2849}}
\affiliation{Department of Physics and Astronomy, University of Tennessee, Knoxville, Tennessee 37996, USA}
\affiliation{Physics Division, Oak Ridge National Laboratory, Oak Ridge, Tennessee 37831, USA}

\date{\today}

\begin{abstract}

The variational quantum eigensolver (VQE) is an algorithm to compute ground and excited state energy of quantum many-body systems. A key component of the algorithm and an active research area is the construction of a parametrized trial wave function  a so\hyp called variational ansatz. The wave function parametrization should be expressive enough, i.e., represent the true eigenstate of a quantum system for some choice of parameter values. On the other hand, it should be trainable, i.e., the number of parameters should not grow exponentially with the size of the system. Here, we apply VQE to the problem of finding ground and excited state energies of the odd-odd nucleus $^{6}$Li. We study the effects of ordering fermionic excitation operators in the unitary coupled clusters ansatz on the VQE algorithm convergence by using only operators preserving the $J_z$ quantum number. The accuracy is improved by two orders of magnitude in the case of descending order. We first compute optimal ansatz parameter values using a classical state\hyp vector simulator with arbitrary measurement accuracy and then use those values to evaluate energy eigenstates of $^{6}$Li on a superconducting quantum chip from IBM. We post-process the results by using error mitigation techniques and are able to reproduce the exact energy with an error of $3.8\%$ and $0.1\%$ for the ground state and for the first excited state of $^{6}$Li, respectively.
\end{abstract}

\maketitle
 
\section{Introduction}
The simulation of static and dynamic properties of quantum many\hyp body systems is a challenging task for classical computers due to the exponential scaling of the Hilbert space. In contrast, quantum computers could be natural devices to solve such problems~\cite{Feynman}, avoiding the exponential scaling. For example,  quantum algorithms such as the quantum phase estimation (QPE) \cite{QPE} 
can perform eigenvalues calculations in polynomial time~\cite{QPE-Lloyd} using future quantum error\hyp corrected hardware. Currently, the circuit depth required to implement them is far greater than that of state\hyp of\hyp the\hyp art noisy intermediates scale quantum (NISQ)~\cite{Preskill2018quantumcomputingin} devices. Nevertheless, NISQ devices have attracted a lot of interest in nuclear physics~\cite{Papenbrock-Deuterium,klco2018pra,LuHH2019,roggero2019,roggero2020,roggero2020b,cervia2021,du2021,Be8-VQE,Qian:2021jxp,LMG,baroni2022,turro2022,Hlatshwayo:2022yqt}. Presently available quantum hardware can be used to compute the ground state energy $E_0$ of a Hamiltonian $H$ by using the variational principle. The variational quantum eigensolver (VQE)~\cite{vqe_original,VQE_Gambetta,VQE-Review} is a hybrid quantum\hyp classical algorithm~\cite{variational_quantum_algorithm} which classically minimizes the expectation value of a trial wave function in the form of a parametrized quantum circuit~\cite{PQC} 
\begin{equation}
    E_0 \leq \frac{\bra{\psi(\theta)}H\ket{\psi(\theta)}}{\braket{\psi(\theta)}}.
\end{equation}

The trainability of the VQE is closely related to the chosen wave function ansatz. It has to be expressive enough to contain the optimal solution, yet simple enough to enable training and to avoid unpleasant effects like the barren plateaus~\cite{Barren_platea_McClean}. Hardware efficient \cite{Panos_excitation_preserving} and physically inspired ans\"atze \cite{magnetogrossi} are popular choices for this task. The former is as shallow as possible in the circuit architecture, with the smallest number of CNOT gates executable on NISQ devices, whereas the latter is built according to properties of the underlying physical system. Although VQE simulations have been widely and successfully used in quantum chemistry~\cite{vqe_original,VQE_Gambetta,UCC_spin_states,UCC-chemistry,Panos_excitation_preserving,magnetogrossi,ADAPT-VQE}, there are fewer applications in nuclear physics \cite{Papenbrock-Deuterium,cervia2021,Be8-VQE,Qian:2021jxp,LMG}. While both fields share many similarities, such as being formulated as non relativistic quantum field theories in second quantization, they differ in many other aspects. For instance, protons and neutrons, the equivalent of $\alpha$ and $\beta$ electrons in quantum chemistry, interact via strong and short-ranged forces, and symmetry breaking, i.e., nuclear deformation and superfluidity, is abundant. This makes it important to reflect this physics in the quantum circuit~\cite{Be8-VQE}.

Starting from the work~\cite{Be8-VQE} on atomic nuclei, we study several training strategies for the convergence of different ans\"atze for the $^{6}$Li nucleus and evaluate our results on superconducting quantum hardware. We note that the papers~\cite{Be8-VQE,LMG} focused on even-even nuclei, which are simpler in structure than the odd-odd nucleus $^6$Li. This makes the problem an interesting step toward VQE applications in nuclear physics.

This paper is organized as follows. We define the theoretical framework in Sec.~\ref{theory},  introduce the model in Sec.~\ref{model}, and present the different ans\"atze used in the present work in Sec.~\ref{ordering}. We present results for the energies of the ground state and first excited states obtained from simulations in Sec.~\ref{statevector} and from superconducting quantum hardware in Sec.~\ref{hardware}.

\section{Theoretical framework}
\label{theory}
We consider a simple shell model where the nucleus $^6$Li is described as a valence proton and neutron added to the inert $^4$He core. In this section, we describe the model space and Hamiltonian, present the unitary coupled-cluster ansatz, and discuss in detail the ordering and implementation of the excitation operators.

\subsection{Model space}
\label{model}

The model space consists of the $0p_{3/2}$ and $0p_{1/2}$ harmonic oscillator orbitals for the neutron and the proton, and we use the Cohen\hyp Kurath interaction \cite{Hamiltonian-COHEN19651}. Our work builds on the recent computation of $^6$He in the same framework~\cite{Be8-VQE} and extends it to a somewhat larger Hilbert space and a somewhat more complicated nucleus. 
In addition to being realistic and non\hyp trivial, our model has the advantage of being simple enough to be run on current NISQ devices. The Hamiltonian can be written in second quantization as
\begin{equation}
    H = \sum_i \epsilon_i \hat{a}_i^\dagger \hat{a}_i +\frac{1}{2}\sum_{ijkl}V_{ijlk} \hat{a}_i^\dagger \hat{a}_j^\dagger \hat{a}_k \hat{a}_l \ .
    \label{hamiltonian}
\end{equation}
Here, $\hat{a}_i^\dagger$ and $\hat{a}_i$ are the creation and annihilation operators , respectively, for a nucleon in the state $|i\rangle$.  The single-particle energies are denoted as  $\epsilon_i$ and two\hyp body matrix elements as $V_{ijkl}$.  
All computed energies are with respect to the ground state energy of the $^4$He core.

We have $|i\rangle = |n=0, l=1, j, j_z, t_z\rangle$, where $n$ and $l$ denote the radial and orbital angular momentum quantum numbers, respectively, $j=1/2, 3/2$ the total spin,  $j_z$ its projection, and $t_z=\pm 1/2$ the isospin projection. Thus, the $p$ shell model space includes six orbitals for the protons and six orbitals for the neutrons, and we need $N=12$ qubits, one per orbital.
The Cohen\hyp Kurath interaction preserves total spin $J$ and total isospin $T$, and their projections $J_z$ and $T_z$. We will exploit that $J_z$ and $T_z$ are conserved in our wave function ansatz.

We convert the shell-model Hamiltonian~(\ref{hamiltonian}) into a qubit Hamiltonian via the Jordan\hyp Wigner~\cite{JW} transformation, i.e., we have the mapping
\begin{equation}
    \hat{a}_i^\dagger = \frac{1}{2}\left(\prod_{j=0}^{i-1}-Z_j\right) (X_i-iY_i),
\end{equation}
\begin{equation}
    \hat{a}_i = \frac{1}{2}\left(\prod_{j=0}^{i-1}-Z_j\right) (X_i+iY_i),
\end{equation}
where $X_i,Y_i$ and $Z_i$ are the Pauli matrices acting on the $i$th qubit.
The Bravyi\hyp Kitaev~\cite{BK} mapping is an alternative to Jordan\hyp Wigner that achieves exponentially shorter Pauli strings in the asymptotic limit. However,  both transformations perform similarly for modest systems sizes~\cite{JW_VS_BK}. As only the Jordan\hyp Wigner mapping enjoys an intuitive translation of the $J_z$ symmetry on the qubit system, we will only consider this mapping in the present work.
Each single-particle state is represented by a qubit where $\ket{0}$ and $\ket{1}$ refer to an empty and an occupied state, respectively. For completeness, we list the different states in Table \ref{tab:label}. 
\setlength{\tabcolsep}{8pt}
\renewcommand{\arraystretch}{1.3}
\begin{table}
\begin{tabular}{|c|ccc|}
\hline
  qubit & \text{ }$j$& $j_z$& $t_z$\\
  \hline
  0 &1/2&$-1/2$ &$-1/2$ \\
  1 &1/2&$+1/2$ &$-1/2$ \\
  2 &3/2&$-3/2$ &$-1/2$ \\
  3 &3/2&$-1/2$ &$-1/2$ \\
  4 &3/2&$+1/2$ &$-1/2$ \\
  5 &3/2&$+3/2$ &$-1/2$ \\
  6 &1/2&$-1/2$ &$+1/2$ \\
  7 &1/2&$+1/2$ &$+1/2$ \\
  8 &3/2&$-3/2$ &$+1/2$ \\
  9 &3/2&$-1/2$ &$+1/2$ \\
      
  10 &3/2&$+1/2$ &$+1/2$ \\
      
  11 &3/2&$+3/2$ &$+1/2$ \\
  \hline
\end{tabular}
  
  \caption{Orbitals represented by the different qubits. Here, $j$ is the total angular momentum, $j_z$ its projection on the $z$ axis and $t_z$ is the third component of the isospin.}

 \label{tab:label}
 \end{table}

Despite the simplicity of our model, the Hamiltonian~(\ref{hamiltonian}) consists of 975 Pauli terms. This large number arises because the short-range nuclear interaction is nonlocal when expressed in the harmonic-oscillator basis, and the number of Pauli terms naively scales as $n^4$, which is reduced by an order of magnitude because of the conservation of spin and isospin.  
Eventually, it could be an advantage to use a lattice formulation~\cite{lee2009} where the short range of the nuclear interaction reduces the number of Pauli terms. On lattices, nuclear Hamiltonians exist that only include next-to-nearest neighbor interactions~\cite{lu2019} and therefor require resources similar to the three-dimensional Hubbard model. At this moment, however, the minimum  $2\times 2\times 2$ lattice requires 32 qubits because of spin and isospin degrees of freedom, and our smaller shell-model space yields more realistic results.

We deal with the large number of Pauli terms by grouping them into 250 sets of qubit-wise commuting operators.
Commuting operators are simultaneously diagonalizable, allowing the computation of the expectation value from the measurements of a single circuit. Additional techniques exist to reduce further the number of circuits. References~\cite{pauli_grouping,Miller_grouping} propose to further group the Pauli operators to include general commutating operators at the cost of appending a circuit with $\mathcal{O}({N^2})$ gates before the measurements. General commutating operators $\mathcal{O}^1,\mathcal{O}^2$ satisfy $[\mathcal{O}^1,\mathcal{O}^2]=0$, whereas qubit\hyp wise commuting operators  $[\mathcal{O}^1_i,\mathcal{O}^2_i]=0$ for all $i$. Reference \cite{efficient_measurement} obtained a cubic reduction by using low\hyp rank factorization. It is even possible to reduce the measurements to a single operator~\cite{POVM}, by using quantum information complete measurements at the cost of a higher number of circuit executions and ancilla qubits. Nonetheless, currently available resources for this work were enough to evaluate the whole Hamiltonian with the qubit\hyp wise commutating grouping. We consequently followed this technique to avoid deeper circuits.

\subsection{The unitary coupled cluster ansatz}
\label{ucc}
The \textit{unitary coupled clusters ansatz (UCC)} is widely used to obtain a correlated ground state from an initial Hartree-Fock  solution $\ket{\psi_0}$ in quantum chemistry and nuclear physics \cite{evangelista2019,UCC-chemistry,UCC_spin_states}. It lets the Hartree-Fock state evolve according to the cluster operator $\hat{T}$. To be compatible with a quantum computing paradigm, the operator has to be unitary. Therefore we choose
\begin{equation}
    \ket{\psi(\bm{\theta})} = e^{i\left(\hat{T}(\bm{\theta})-\hat{T}^\dagger(\bm{\theta})\right)}\ket{\psi_0}.
\end{equation}
$\hat{T}$ can be decomposed into singles, i.e., one-particle--one-hole (1p-1h), doubles (2p-2h), ...,  excitation operators of the following form:
\begin{equation}
    \hat{T} = \hat{T}_1 + \hat{T}_2 + \dots
\end{equation}
with 
\begin{equation}
    \hat{T}_1 = \sum_{i\in \text{virt}; \alpha \in \text{occ}} \theta_i^\alpha \hat{a}_i^\dagger \hat{a}_\alpha
\end{equation}
and
\begin{equation}
    \hat{T}_2 = \sum_{i,j\in \text{virt}; \alpha, \beta \in \text{occ}} \theta_{ij}^{\alpha \beta} \hat{a}_i^\dagger \hat{a}_j^\dagger \hat{a}_\alpha \hat{a}_\beta.
\end{equation}
In the above definitions, the Latin indices run over virtual (empty) states and the Greek over occupied states of the initial state. The cluster operator drives occupied orbitals to empty ones.
To respect symmetries and reduce the number of terms, we only considered excitation operators with a total angular-momentum projection $J_z=0$. The Jordan\hyp Wigner mapping is then used to transform the unitary cluster ansatz into a qubit operator with trainable parameters $\bm{\theta}$. The UCC ansatz is finally implemented using one step of the first order Trotter formula.

The \textit{initial state} $\ket{\psi_0}$ is usually chosen as the Hartree\hyp Fock solution. However, it is often sufficient to lie close enough to the actual ground state. For instance, the $^6$Li ground state has spin $J=1$ and therefore $J_z=-1, 0$, or $1$. So, any product state with this configuration, e.g., $\ket{1} \otimes \ket{6}$ or  $\ket{0} \otimes \ket{7}$ should converge to the ground state. Moreover, this observation can help us find the first excited state (with spin $J=3$), which lies in the subspace with a total $J_z$  of $-3$, $-2$, 2, or 3 orthogonal to the ground state. This observation provides a particular advantage over other methods in finding excited states with the VQE, such as the iterative constrained optimization~\cite{Excited-states}, the discriminative VQE~\cite{DVQE}, or those based on the quantum equation of motion~\cite{Ollitrault_excited_state}. These techniques require additional quantum or classical resources and rely on the accuracy of the prepared ground state, therefore suffering from the error amplification phenomenon. On the other hand, enforcing the ansatz to stay in a particular region of the Hilbert space by choosing the right quantum numbers, produces stable and accurate solutions which are easy to obtain,  when applicable.

\subsection{Excitation ordering}
\label{ordering}
In the following, we describe different strategies to study the convergence of the variational method.
The ordering of the excitation operators impacts the training landscape and the convergence behavior. Hence, an ansatz may quickly converge while another remains trapped in a local minimum. We observed this in our work by trying different ordering strategies.

\textit{Shuffling} is a strategy that consists of choosing the best sorting over multiple runs with a random shuffling. It quickly becomes prohibitive to explore the shuffled space when enlarging the system size, but is interesting to consider since the ordering has a non\hyp trivial effect on the optimization procedure. We will refer to this strategy as \emph{best shuffle} throughout this paper.

\textit{Ordering} represents a second option, where we proceed to order the operators by their absolute magnitude of the corresponding term in the Hamiltonian. Hence, the singles excitation refers to the corresponding single-particle energy $\epsilon_i$ while the doubles excitation $\hat{a}_i^\dagger \hat{a}_j^\dagger \hat{a}_\alpha \hat{a}_\beta$ refers to the two\hyp body term $V_{ij\beta \alpha}$. The considered Hamiltonian permits only to apply this ordering on singles and doubles terms, but more complex models could be considered to order 3p\hyp 3h or 4p\hyp 4h excitation as well. This approach, similar to the QDrift~\cite{QDrift_campbell} algorithm for time evolution (which chooses the terms randomly to be evolved according to their relative magnitude), orders the excitation operator in descending order of magnitude such that the most important ones are placed at the beginning. We observed that this technique, which we will refer to as \emph{ordered} UCCSD, is the most promising ansatz among the ones considered in this work. Moreover, when coupled to a layerwise learning scheme, it achieves arbitrary accuracy in a polynomial number of optimization steps.

\textit{Adaptive derivative\hyp assembled problem\hyp tailored (ADAPT)\hyp VQE} is another efficient strategy to adaptively order the operators with respect to the magnitude of their gradient. ADAPT\hyp VQE \cite{ADAPT-VQE} constructs the ansatz by picking from a pool of operators $\{\hat{\tau}_0, \dots, \hat{\tau}_n\}$ the one which has the most impact on the expectation value, namely the one with the largest gradient magnitude
\begin{equation}
   \left| \frac{\partial E}{\partial \theta_i}\right|_{\theta_i=0} = |\bra{\psi}[H,\hat{\tau}_i]\ket{\psi}|.
\end{equation}
The chosen operator is recursively added to the current ansatz, leading to 
\begin{equation}
    \ket{\psi(\bm{\theta})} =e^{-i\theta_l\tau_l} e^{-i\theta_{l-1}\tau_{l-1}}\dots e^{-i\theta_0 \tau_0} \ket{\psi_0},
\end{equation} 
after adding $l$ operators. We set $\theta_l=0$ to allow a smooth transition between the architecture's update. The picking action is followed by $k$ optimization steps, and it is repeated until convergence is reached. We note that in the original proposition, the circuits are optimized until convergence, before picking a new term. However in our experience, it may be beneficial to apply early stopping, after $k$ steps to avoid being trapped in a local minima. Hence, we compared $k=10$ to $k=100$, and only $k=10$ was able to reach an error ratio below $1\%$. It generally leads to accurate solutions with minimal depth. The computation of the gradients of all the operators in the pool, which is time-consuming, can in principle be performed in parallel.

\textcite{ADAPT-VQE} demonstrated with numerical experiments that ADAPT\hyp VQE is superior to random or lexical ordering of the excitation operators in terms of convergence and circuit depth. However, our study suggests that reducing the operator pool using symmetries and ordering with respect to their magnitude achieves quicker convergence. Studies of the Lipkin-Meshkov-Glick model~\cite{LMG} showed that the number of operators needed to achieve $1\%$ accuracy increases linearly with the number of valence neutrons. However, this behavior has only been simulated within nuclei with an even number of valence neutrons and without valence protons: it remains an open question whether this result also holds with neutron\hyp proton interactions.

Finally, we consider \textit{layerwise learning}, a technique initially proposed to mitigate barren plateaus in quantum machine learning \cite{layerweise}. The idea is to consider $m$ singles terms first, perform $k$ optimization steps, add $m$ new singles terms, and continue until all singles terms have been used before moving to higher\hyp order interactions. In an ordered approach, the first $m$ operators added to the ansatz are instead chosen according to the selected ordering. We choose $m=1$ for the remaining of this paper.

\subsection{Hardware efficient ansatz}

Because of the Jordan\hyp Wigner mapping, fermionic excitation operators act on $\mathcal{O}(N)$ qubits. They are therefore expensive to be implemented on NISQ devices due to the increased connectivity requirement and the consequent increase in the number of CNOT and Swap gates needed after circuit transpilation. A simple and alternative way to reduce this expense is to consider qubit\hyp based excitation (QBE) \cite{QBE_original,QBE} operators. QBE efficiently implements the excitation operators on $\mathcal{O}(1)$ qubits by neglecting the $Z$ terms in the Jordan\hyp Wigner mapping. Essentially, creation operators are mapped to
\begin{equation}
    \hat{a}_i^\dagger = \frac{1}{2} (X_i-iY_i),
\end{equation}
and annihilation operators to 
\begin{equation}
    \hat{a}_i = \frac{1}{2} (X_i+iY_i).
\end{equation}
The difference between the Jordan\hyp Wigner mapping is that the resulting operator will not respect fermionic anti commutation relations, which are enforced by the product of Pauli $Z$ matrices. Single excitation operators between qubits $i$ and $j$ read
\begin{equation}
    U_{ij}(\theta) = \text{exp}\left[i\frac{\theta}{2}(X_iY_j-Y_iX_j)\right]
\end{equation}
and double excitation operator between qubits $i,j,k$ and $l$ are
\begin{align}
\begin{split}
    U_{ijkl}(\theta) = \text{exp}\Big{[}i\frac{\theta}{8}(&X_iY_jX_kX_l+Y_iX_jX_kX_l \\ +&Y_iY_jY_kX_l+Y_iY_jX_kY_l \\
    -&X_iX_jY_kX_l-X_iX_jX_kY_l \\
    -&Y_iX_jY_kY_l-X_iY_jY_kY_l)\Big{]}.
\end{split}
\end{align}
Even if QBE-UCC ans\"atze do not respect the fermionic anti commutation relations, they show a comparable efficiency for ground state calculations. Those ans\"atze are hardware efficient as they act on a fixed number of qubits (two for the singles, four for the doubles, and $2^k$ for the $k$th excitation operators). The exact circuit formulation can be found in the original paper~\cite{QBE_original}.
 
Finally, we also considered an efficient excitation\hyp preserving ansatz (EPA), such as the one proposed in \cite{Panos_excitation_preserving,symmetry_economou} for quantum chemistry. These ans\"atze are based on gates preserving the number of occupied orbitals. Moreover, time\hyp reversal symmetry can lead to further simplifications. They have the advantage of using fewer CNOT gates resulting in a more shallow circuit, an advantage for near\hyp time devices. However, they cannot respect the total $J_z$ symmetry as they act on the protons and neutrons separately. In our investigations, this led to circuits suffering from barren plateaus~\cite{Barren_platea_McClean} which are expected in generic circuits using a global cost function~\cite{local_cost_function}, such as the expectation value of the Hamiltonian. We remark that the gradient vanishes from the beginning, and changing the number of layers, optimizer, learning rate, parameters initialization, and even using an automatic differentiation framework to compute the gradient did not permit us to train the ansatz. This observation suggests that symmetries play a non\hyp negligible role in nuclear structure calculations since it is the significant difference between UCC based ans\"atze and excitation preserving ones. More details about the construction of this type of ans\"atze can be found in the Appendix .
 
\section{Results}
 
In the following, we present the results obtained with the different circuit architectures discussed in Sec.~\ref{ordering}. The investigations are performed on a state\hyp vector simulator and the hardware\hyp friendly QBE\hyp UCCSD ansatz is evaluated on a real quantum processor. 

State\hyp vector simulations allow one to probe the potential of this approach under ideal conditions, such as using exponentially many shots or without noise. UCC ans\"atze are notoriously deep, and the noise heavily deteriorates the outcome, even when using error mitigation techniques. We address these difficulties in the following, showing the results step by step.
 
\subsection{Optimization}
For the optimization, we use the simultaneous perturbation stochastic approximation (SPSA)~\cite{SPSA} optimizer with a fixed number of iterations. SPSA efficiently approximates the gradient with only two circuit evaluations by shifting the parameters in two random directions. The learning rate \texttt{lr} = 0.1, is halved at every 25 iterations until \texttt{lr} = 0.001 to ensure a fast convergence at the beginning and avoid oscillations at the end. Looking at realistic experiments, the stochastic nature of SPSA makes it resilient to the statistical noise coming from the finite number of measurements, making it appealing for quantum devices. All the initial parameters, except the first one, are set to zero at the beginning of the optimization in an attempt to mitigate barren plateaus~\cite{initialization}, while the first is chosen at random between 0 and $2\pi$, but fixed for the different experiments. We remark that the value of the first parameter has a negligible effect on the convergence.

A quantum natural variant of the SPSA optimizer using the geometry of the Hilbert space has been recently proposed by ~\textcite{QNSPSA}. It uses six circuit evaluations to approximate the Hessian (which can be used to compute the quantum natural gradient) and significantly improves the optimization efficiency of quantum circuits. In the present work, the effect of the quantum natural gradient is mainly appreciated on hardware\hyp friendly ansatz such as QBE-UCC.

We did not consider any other optimization methods, such as the gradient\hyp based ADAM \cite{adam} or the gradient\hyp free COBYLA \cite{cobyla} optimizer. Hence, SPSA only requires two sampling execution per step and is resilient to statistical noise due to its stochastic nature, making it one of the most suitable optimizer for NISQ devices. Moreover, we did not need to use more expensive techniques since it achieved an exponential convergence for this particular case. However, we did try ADAM and COBYLA for the EPA since they were suffering from BP. Nevertheless, it did not change the optimization landscape which remained flat, as argued in Ref. \cite{Arrasmith2021effectofbarren}.

\subsection{State-vector simulations}
\label{statevector}

The gate\hyp based quantum circuits used in this section are built using the open\hyp source framework \href{https://qiskit.org/documentation/nature/}{\texttt{qiskit-nature}} \cite{Qiskit} and are run on pennylane~\cite{pennylane} using the C++ \href{https://pennylane-lightning.readthedocs.io/en/latest/devices.html}{\texttt{lightning.qubit}} plugin. 

\begin{figure}
    \centering
    \includegraphics[scale=0.53]{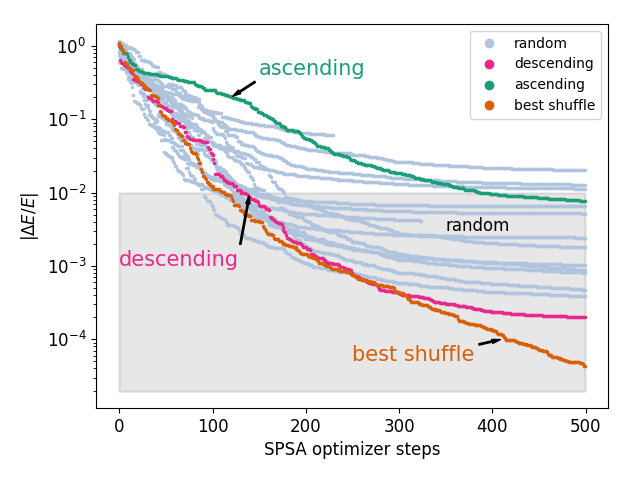}
    \caption{Training curve in a semilog scale for fermionic UCC ansatz with different ordering. The \emph{best shuffle} curve is taken among 20 independent runs. The grey area corresponds to the 1\% margin, which is acceptable in most applications.}
    \label{fig:6Li_oredring}
\end{figure}

We first assess the effect of \textit{ordering} on the fermionic\hyp UCC ansatz starting from the initial state $\ket{2}\otimes \ket{11}$. This state has $J_z = 0$, and has the largest operator pool on which we perform $500$ optimization steps. The optimization curve, which shows the error ratio 
 \begin{equation}
     \left| \frac{\Delta E}{E}\right|:=  \left|\frac{E_{\text{VQE}}-E_{\text{exact}}}{E_{\text{exact}}}\right|,
 \end{equation} 
for different ordering is shown in Fig.~\ref{fig:6Li_oredring}. We observe that the descending ordering (pink) strategy leads to fast convergence while the ascending ordering (green) strategy converges slowly. Thus, most important operators should be placed first. We also note that a favorable convergence trend is also given by the \emph{best shuffle} curve (orange), which is taken among 20 independent runs, and by some random run combinations, for which the relative differences are not easily interpreted.
 \\
 
\paragraph{Ground state calculation}
We now compare the different ans\"atze presented in Sec.~\ref{ordering} to prepare the ground state. For the fermionic\hyp UCC ansatz, we start again from $\ket{2}\otimes \ket{11}$ state, and we train with the SPSA optimizer. In the iterative approaches (ADAPT-VQE, Layerwise Learning), $k=10$ iterations are performed between each architecture update. This choice has shown to be a good trade\hyp off between a slow convergence (for large $k$) and deep circuits (for small $k$). For the QBE-UCC ansatz, more favorable results were obtained using $\ket{0}\otimes \ket{7}$ as initial state, which also has the smallest operator pools and is consequently better suited for noisy devices. 

\begin{figure}
     \centering
     \includegraphics[scale=0.53]{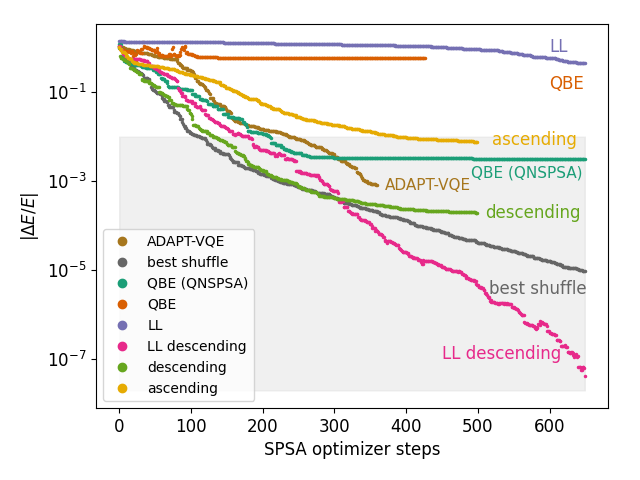}
     \caption{Training curve in a semilog scale for all different ans\"atze with different training strategies. The grey area corresponds to the 1\% margin, which is acceptable in most applications.}
     \label{fig:loss_6Li}
 \end{figure}

The learning curves are shown in Fig.~\ref{fig:loss_6Li} and we observe that descending ordering strategies are among the fastest and more accurate ones, the best being the layerwise learning with descending ordering (pink). Interestingly, the ADAPT-VQE (brown) approach does not perform as well as the former. We suspect that the gradient evaluated at $\theta = 0$  does not contain enough information to obtain the optimal solution. On the one hand, the circuit at the beginning is too shallow, which explains the slow convergence curve. On the other hand, the algorithm mainly picks the same operators which may prevent convergence to the optimal solution. Hence, it only used half of the available operators before becoming too deep to be trained efficiently.

The quantum natural SPSA optimizer significantly improves the optimization of the QBE\hyp UCC ansatz, compared to the standard SPSA. This can be seen by comparing the QBE (QNSPSA) curve (dark green) with the QBE (orange) one. The descending ordered UCC ansatz achieves an exponentially fast convergence, that can be deduced from the linear behavior on a log scale, and the descending layerwise learning strategy reaches arbitrary accuracy, since it approaches the 16 digits numerical precision of the ground truth exponentially fast. 
 \\
 
\paragraph{First excited state}
As pointed out in Sec.~\ref{ucc}, the choice of the initial state allows us to find the first excited state. Hence, an initial state with $J_z = -3,\, -2,\; 2,$ or $3$ will remain in a subspace is orthogonal to the ground state. For instance, the state $\ket{1} \otimes \ket{11}$ has $J_z=2$ and it is therefore a possible candidate. The result obtained in this case, for a fermionic UCC ansatz optimized on a statevector simulator, achieves an error ratio of $10^{-11}$.

\subsection{Hardware}
\label{hardware}
We evaluate the most hardware\hyp efficient ansatz, i.e.,  QBE\hyp UCC, previously trained on the statevector simulator, on a superconducting chip from IBM. Gate\hyp based quantum circuits, ran on the cloud using the IBM Quantum Lab, are transpiled onto the hardware topology by using the SWAP-based BidiREctional (SABRE) heuristic search algorithm~\cite{sabre}. Multiple runs are performed to select the circuit which minimizes the total number of CNOT gates needed. The SABRE algorithm enables a 50\% CNOT reduction compared to a naive approach for a total of 209 CNOTs. Measurement error mitigation is performed efficiently, as proposed in Ref.~\cite{MEM}, by individually inverting the error matrices 
\begin{equation}
S_k=
\begin{pmatrix}
P^{(k)}_{0,0} &P^{(k)}_{0,1} \\P^{(k)}_{1,0} &P^{(k)}_{1,1}
\end{pmatrix}.
\end{equation}
Here, $P^{(k)}_{i,j}$ is the probability of the $k$th qubit to be in state $j\in\{0,1\}$ while measured in state $i\in\{0,1\}$. While this only corrects the uncorrelated readout errors, it is argued in Ref.~\cite{MEM} that they are the predominant ones, making it a useful tool for measurement error mitigation for large number of qubits.

Regarding CNOT errors, the zero\hyp noise extrapolation  \cite{ZNE,Gambetta-Richardson_extrapolation,error-mitigation} represents a powerful mitigation technique by artificially stretching the noise to be extrapolated to the noise\hyp less regime. However, the structure of the considered Hamiltonian amplified the effect of CNOT errors considerably and prevented us from using this strategy. Hence,  states with a wrong number of occupied orbitals belong to different nuclei, which can have much lower energy. We observed a discrepancy of almost 300\% by stretching the noise with a factor two and did not investigate zero\hyp noise extrapolation further.

Our tests are executed on the IBM Quantum 27 qubits architecture \emph{ibmq\_mumbai} and repeated ten times for the ground state and five for the first excited state, using $8092$ shots each. The results are reported in Table \ref{tab:hardware}, alongside the number of parameters and CNOT gates after transpilation. We observe that the energy is reproduced up to $3.81\%$ and $0.12\%$ accuracy for the ground and first excited state, respectively. Both lie within one standard deviation confidence interval. Moreover, the standard deviation for the ground state is ten times smaller than the energy gap with the first excited state, which accentuates the accuracy of our results. We remark that measurement error mitigation increases the accuracy by more than 10\%, making it appealing for readout error mitigation in quantum circuits with a large number of qubits.

\begin{center}
\setlength{\tabcolsep}{8pt}
\renewcommand{\arraystretch}{1.3}
\begin{table*}
\begin{tabular}{|c|cccccc|}
\hline 
hardware &No. parameters & No. CNOT& mean & st. deviation & exact & error ratio\\
  \hline
 ibmq\_mumbai raw (gs) & 9& 209& -6.27 &0.269& -5.529 &13.36\% \\
 ibmq\_mumbai mitigated (gs) & 9& 209 &-5.319 &0.24& -5.529 &3.81\% \\
  ibmq\_mumbai raw (1st es) & 3& 41& -2.907 &0.87& -3.420 &14.97\% \\
 ibmq\_mumbai mitigated (1st es) & 3& 41 & -3.424 &0.08& -3.420 &0.12\% \\
 \hline 
\end{tabular}
  
  \caption{Hardware results of the QBE-UCC ansatz for the ground state (gs) and first excited state (1st es), alongside the number of parameters and CNOT gates after transpilation. The exact result, obtained with exact diagonalization,  are reproduced up to one standard deviation.}
 
 \label{tab:hardware}
 \end{table*}
\end{center}

\section{Conclusions}
We performed shell-model quantum-computations of the nucleus $^6$Li, composed of a frozen $^4$He core and two valence nucleons.
We studied the effect of the ordering of excitation operators in unitary coupled clusters type ans\"atze for the variational quantum eigensolver. We empirically observed that the ordering strongly affects the learning curve and that arranging in descending order of magnitude with respect to the Hamiltonian leads to a better convergence behavior than random ordering or ADAPT\hyp VQE. Hence, operators with high magnitude have more importance in the system's description, which should be reflected in the ansatz construction. Moreover, adopting a layerwise learning scheme, where the operators are iteratively added to the circuit, has shown an accuracy of the order of $10^{-7}$.
By choosing an initial state with a suitable $J_z$ quantum number, we were also able to compute the energy of the first excited state with a precision of $10^{-11}$.

Finally, we evaluated the qubit based excitation-UCC (QBE-UCC), which neglects the fermionic anti commutation relation to reduce the number of CNOT and SWAP gates needed. We performed, for the first time to our knowledge, these calculations on a real quantum device, a 27 qubits machine (\emph{ibmq\_mumbai}), and we were able to reproduce the exact ground state and first excited state energy up to one standard deviation.

The number of nuclear states grows factorially with the number of valence nucleons, making the scaling of VQE applications impractical. Even if the numbers of singles and doubles excitation operators seem to grow linearly~\cite{LMG}, it may be necessary to use triples and quadruples excitation operators as well. Reference~\cite{Be8-VQE} demonstrated that quadruple operators acting on all valence nucleons were necessary in a UCC ansatz for a $^8$Be nucleus, composed of two protons and two neutrons in the $p$ shell, and achieved 1\% error ratio on statevector simulations with 118 parameters. This motivates symmetry considerations to reduce the number of operators, in order to prevent deep ans\"atze, which are not easily trainable, while keeping all the operators needed to reproduce the exact energy. This will be the focus of future research in this direction.

\begin{acknowledgments}
We thank Zhonghao Sun for helping with the nuclear matrix elements. This work was supported by CERN Quantum Technology Initiative, the U.S. Department of Energy, Office of
Science, Office of Nuclear Physics, under Award Nos.~DE-FG02-96ER40963 and DE-SC0021642, and by the Quantum Science Center, a National Quantum Information Science Research Center of the U.S. Department of Energy. Oak Ridge National Laboratory is
supported by the Office of Science of the U.S. Department of Energy under Contract No. DE-AC05-00OR22725.
Access to the IBM Quantum Services was obtained through the IBM Quantum Hub at CERN. The views expressed are those of the authors and do not reflect the official policy or position of IBM, the IBM~Q team or AWS.

\end{acknowledgments}
\bibliography{bibliography,master}

\appendix 
\section{Excitation-preserving ansatz}
\label{EPA}
In this appendix, we recall the construction of the excitation\hyp preserving ansatz (EPA), as presented in \cite{Panos_excitation_preserving, symmetry_economou}. As the name suggests, these types of circuits preserve the total number of excited orbitals present in the initial state. In the context of this work, this ensures that the states obtained with the VQE represent the same nuclei, which are defined in terms of number of protons and neutrons. The building blocks of the EPA  $U(\theta,\phi)$, are two-qubit gates which are themselves excitation\hyp preserving and read
\begin{equation}
    U(\theta,\phi) = \begin{pmatrix}
    1&0&0&0\\ 0&\cos{(\theta)}&e^{i\phi}\sin{(\theta)}&0 \\
    0&e^{-i\phi}\sin{(\theta)}&-\cos{(\theta)}&0 \\
    0&0&0&1
    \end{pmatrix}
\end{equation}

\begin{figure}
    \centering
    \includegraphics[scale=0.48]{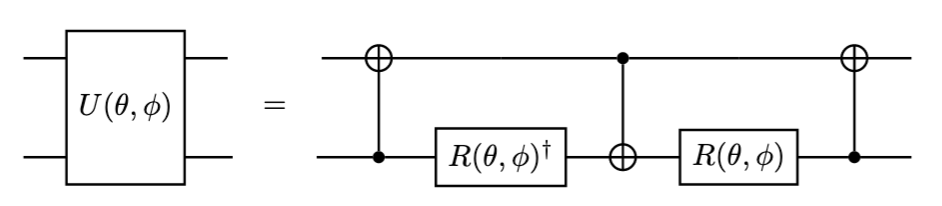}
    \caption{U$(\theta,\phi)$ gate decomposition. Decomposition of $U(\theta,\phi$) in terms of elementary gates where $R(\theta,\phi)=R_z(\phi+\pi)R_y(\theta+\pi/2)$ and $R_z(\theta) = e^{-i\theta Z/2}$, $R_y(\theta) = e^{-i\theta Y/2}$.}
    \label{fig:U}
\end{figure}

This is a valid choice since this matrix maps 
\begin{align}
    \ket{00} &\mapsto \ket{00},\\
    \ket{01} &\mapsto \cos{(\theta)}\ket{01}+e^{i\phi}\sin{(\theta)}\ket{10},\\
    \ket{10}&\mapsto e^{-i\phi}\sin{(\theta)} \ket{01} -\cos{(\theta)}\ket{10}, \\
     \ket{11}&\mapsto \ket{11}.
\end{align}
This gate can be efficiently implemented on a gate\hyp based quantum circuit as shown in \figref{fig:U}.

\begin{figure}
    \centering
    \includegraphics[scale=0.4]{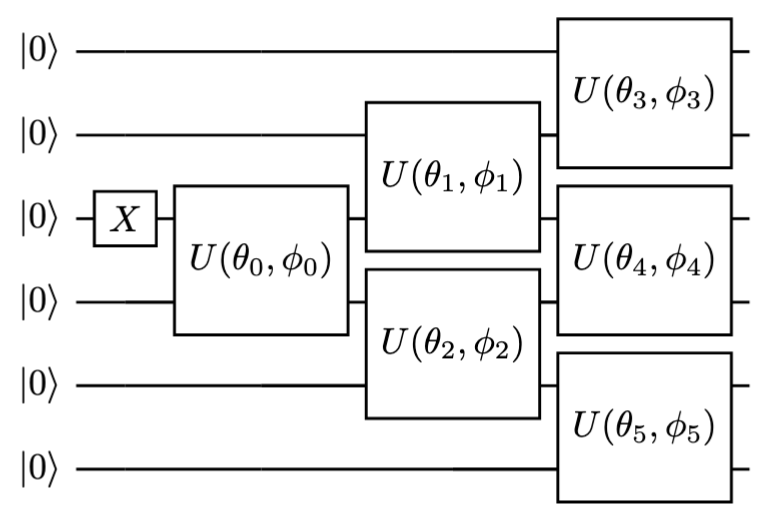}
    \caption{Excitation\hyp preserving ansatz for one single register (e.g. for the protons). The circuit starts from the initial state $\ket{001000}$ and is built with the excitation preserving gate $U(\theta,\phi)$ in a pyramidal and efficient way. }
    \label{fig:SP}
\end{figure}

The ansatz is then built on top of an initial product state with the correct number of excited orbitals $p$, using $\binom{n}{p}$ $U(\theta,\phi)$ gates, where $n=6$ is the total number of orbitals in each register. This precise number of gates \cite{symmetry_economou} is chosen such that the ansatz is maximally expressive, while having the minimal amount of parameters. It is preferable, because of the limited connectivity of NISQ devices, to act only neighboring qubits, starting from the occupied orbital in a pyramidal manner. The $^6$Li nuclei consists of one proton and one neutron. Therefore $p=1$ in both the proton and neutron registers and we choose as initial state $\ket{001000}$ for both of them, as illustrated in \figref{fig:SP}.
The main bottleneck of the EPA is its inability to take into account global symmetries, like the conservation of the total angular momentum $J_z$. Hence, the proton and neutron  registers can not be entangled, since it would enlarge the Hilbert space of the final state, which could have two neutrons and zero protons ($^6$He), or zero neutrons and two protons ($^6$Be).  In practice, we run into barren plateaus early on in the training phase \cite{Barren_platea_McClean}, and  gradient\hyp based (ADAM \cite{adam}, stochastic gradient descent), gradient\hyp free (COBYLA \cite{cobyla}, SPSA \cite{SPSA}) optimizers or layerwise learning \cite{layerweise} strategy did not permit us to avoid them. 
We note that time reversal symmetry can be accounted for by setting $\phi_i=0$ $\forall i $. Nevertheless, this additional symmetry did not change the general behavior of the EPA's optimization.

\end{document}